\documentclass[proof]{WileyASNA-v1}

\articletype{Refereed Conference Paper}%

\received{13 September 2024}
\revised{23 October 2024}
\accepted{28 October 2024}

\begin{document}

\newcommand{\vol}[2]{$\,$\bf #1\rm , #2.}                 
\def\etal{{et al.~}}
\def\ie{{\it i.e.}}
\def\eg{{\it e.g.}}
\def\teq#1{$\, #1\,$}
\def\mgb{M.~G. Baring}

{\catcode`\@=11                                                  
   \gdef\SchlangeUnter#1#2{\lower2pt\vbox{\baselineskip 0pt\lineskip0pt    
   \ialign{$\m@th#1\hfil##\hfil$\crcr#2\crcr\sim\crcr}}}}           
\def\gtrsim{\mathrel{\mathpalette\SchlangeUnter>}}               
\def\lesssim{\mathrel{\mathpalette\SchlangeUnter<}}
\def\rns{R_{\hbox{\sixrm NS}}}
\def\subrm#1{\hbox{\sevenrm #1}}                                 
%
%
\font\fiverm=cmr5       \font\fivebf=cmbx5      \font\sevenrm=cmr7     
\font\sevenbf=cmbx7     \font\eightrm=cmr8      \font\eighti=cmmi8     
\font\eightsy=cmsy8     \font\eightbf=cmbx8     \font\eighttt=cmtt8    
\font\eightit=cmti8     \font\eightsl=cmsl8     \font\sixrm=cmr6       
\font\sixi=cmmi6        \font\sixsy=cmsy6       \font\sixbf=cmbx6      
%
%
%
\def\pr{Phys. Rev.}                             
\def\aapl{Astron. Astr. (Lett.)}                
\def\angeo{Ann. Geophys.}                       
\def\app{Astroparticle Phys.}                   
\def\apss{Astr. Space Sci.}                     
\def\asr{Adv. Space Res.}                       
\def\jetp{Sov. Phys. JETP}                      
\def\phfl{Phys. Fluids}                         
\def\prl{Phys. Rev. Lett.}                      
\def\rmp{Rev. Mod. Phys.}                       
\def\ssr{Space Sci. Rev.}                       
\def\edithere#1{\textcolor{red}{\bf #1}}  
\def\actionitem#1{\textcolor{blue}{\bf #1}}  
\def\editherealso#1{\textcolor{green}{\bf #1}}  
\def\citeblue#1{\textcolor{blue}{[#1]}}
\def\hruleseparator{\vskip 10pt \centerline{\vbox{\hrule height 1.7pt width 60pt}}}

\def\etalc{\it et al., \rm\hskip 0.1em }
\def\taupp{\tau_{\gamma\gamma}}
\def\sigpp{\sigma_{\gamma\gamma}}
\def\sigt{\sigma_{\hbox{\sixrm T}}}
\def\taut{\tau_{\hbox{\sixrm T}}}
\def\pprod{\gamma\gamma\to e^+e^-}
\def\erg{\varepsilon}
\def\eperp{\vphantom{(}\erg_{\perp}}
\def\rns{R_{\hbox{\sevenrm NS}}}                                
\def\mns{M_{\hbox{\sevenrm NS}}}                                
\def\emax{\erg_{\hbox{\sixrm MAX}}} 
\def\dlum{d_{\hbox{\fiverm L}}}
\def\lambar{\lambda\llap {--}}
\def\fsc{\alpha_{\hbox{\sevenrm f}}}                                
\def\dover#1#2{\hbox{${{\displaystyle#1 \vphantom{(} }\over{
   \displaystyle #2 \vphantom{(} }}$}}                
\def\split{\gamma\to\gamma\gamma}
\def\lambdac{\lambda_{\hbox{\sevenrm c}}}  
\def\ecyc{\omega_{\hbox{\fiverm B}}}
\def\gamb{\gamma_{\hbox{\fiverm B}}}   
\def\gammat{\gamma_{\hbox{\fiverm T}}}   
\def\gammin{\gamma_{\hbox{\sixrm min}}}   
\def\ThetaBnone{\Theta_{\hbox{\sixrm Bn1}}} 
\def\Machson{{\cal M}_{\hbox{\fiverm S}}}   
\def\Chandra{{\it Chandra}}  
\def\thetaB{\theta_{\hbox{\fiverm B}}}
\def\muB{\boldsymbol{\mu}_{\rm B}}
\def\IXPE{{\it IXPE}}

\title{Pulsed and Polarized X-ray Emission from Neutron Star Surfaces} 

\author[1]{Matthew G. Baring*}

\author[1]{Hoa Dinh Thi}

\author[2,3]{George Younes}

\author[4]{Kun Hu}

\authormark{Matthew G. Baring \textsc{et al}}

\address[1]{\orgdiv{Department of Physics and Astronomy}, \orgname{Rice University}, \orgaddress{Houston, \state{Texas}, \country{U.S.A.}}}

\address[2]{\orgdiv{Astrophysics Science Division}, \orgname{NASA's Goddard Space Flight Center}, \orgaddress{Greenbelt, \state{Maryland}, \country{U.S.A.}}}

\address[3]{\orgdiv{CRESST}, \orgname{University of Maryland Baltimore County}, \orgaddress{Baltimore, \state{Maryland}, \country{U.S.A.}}}

\address[4]{\orgdiv{Physics Department, McDonnell Center for the Space Sciences}, \orgname{Washington University in St. Louis}, \orgaddress{St. Louis, \state{Missouri}, \country{U.S.A.}}}

\corres{*Matthew G. Baring, Department of Physics and Astronomy, Rice University, 6100 Main Street, Houston, TX 77005-1892, U.S.A. \email{baring@rice.edu}}


\abstract{
The intense magnetic fields of neutron stars naturally lead to strong
anisotropy and polarization of radiation emanating from their surfaces,
both being sensitive to the hot spot position on the surface.
Accordingly, pulse phase-resolved intensities and polarizations depend on the
angle between the magnetic and spin axes and the observer’s viewing
direction. In this paper, results are presented from a Monte Carlo
simulation of neutron star atmospheres that uses a complex electric
field vector formalism to treat polarized radiative transfer due to
magnetic Thomson scattering.  General relativistic
influences on the propagation of light from the stellar
surface to a distant observer are taken into account.  The paper outlines a
range of theoretical predictions for pulse profiles at different X-ray
energies, focusing on magnetars and also neutron stars of lower
magnetization. By comparing these models with observed intensity and polarization 
pulse profiles for the magnetar 1RXS J1708-40, and the light curve
for the pulsar PSR J0821-4300, constraints on the stellar geometry angles 
and the size of putative polar cap hot spots are obtained.}




\maketitle

\footnotetext{\textbf{Abbreviations:} QED: quantum electrodynamics}

\section{Introduction}\label{sec:intro}

Neutron stars are the most magnetic and among the most luminous 
compact objects in the Milky Way.  Their pulsed steady X-ray emission 
enables the estimation of their surface field strengths, which for 
{\bf magnetars} normally exceeds \teq{10^{14}}Gauss.
Soft X-ray pulsation is the telltale signature of non-uniform emission 
across a neutron star surface.  This feature has been exploited by several 
groups to probe the size of the hot emission region(s), 
the value of the star's magnetic inclination angle \teq{\alpha} to 
rotation axis \teq{\boldsymbol{\Omega}} in a dipolar 
field morphology, and the observer's viewing direction angle \teq{\zeta} relative to 
\teq{\boldsymbol{\Omega}}.  For instance, \cite{Albano-2010-ApJ}
fit the pulse profiles of the magnetar
CXOU~J164710.2-455216 (finding \teq{\alpha\sim 80^{\circ}} 
and \teq{\zeta\sim 25^{\circ}}), employing blackbody emission
from the whole surface, and including gravitational light bending.  With a similar construct,
\cite{Guillot-2015-MNRAS} found that a very small polar region with a 
temperature a factor of 6 larger than the equatorial zone is required to fit the
single-peaked soft X-ray profile for the ``low field magnetar'' SGR~0418+572.  
For a non-magnetar example, \cite{Gotthelf-2010-ApJ} modeled the 
energy-dependent light curves of the pulsar PSR J0821-4300 in the supernova remnant Puppis A, finding 
that their best solution was for a quasi-orthogonal rotator viewed almost down the spin axis.
Omitted from these and other similar studies are detailed treatments of local radiation anisotropy, which varies 
in atmospheres as the direction and magnitude of {\bf B} changes across the
stellar surface. 

\footnotetext{\textbf{Abbreviations:} \IXPE: Imaging X-ray Polarimetry Explorer}

\begin{figure*}[t]
\centerline{\includegraphics[width=240pt]{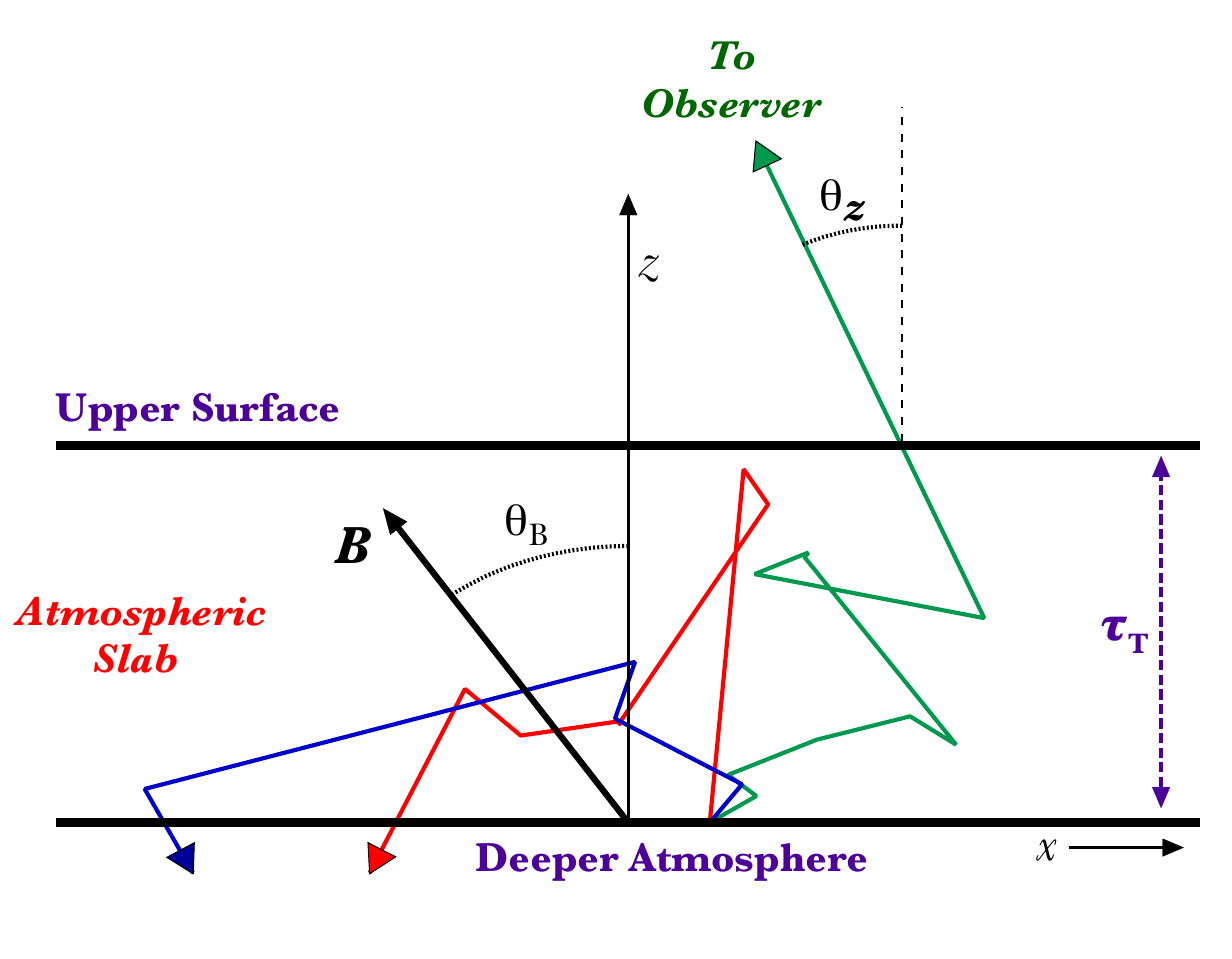}
\includegraphics[width=250pt]{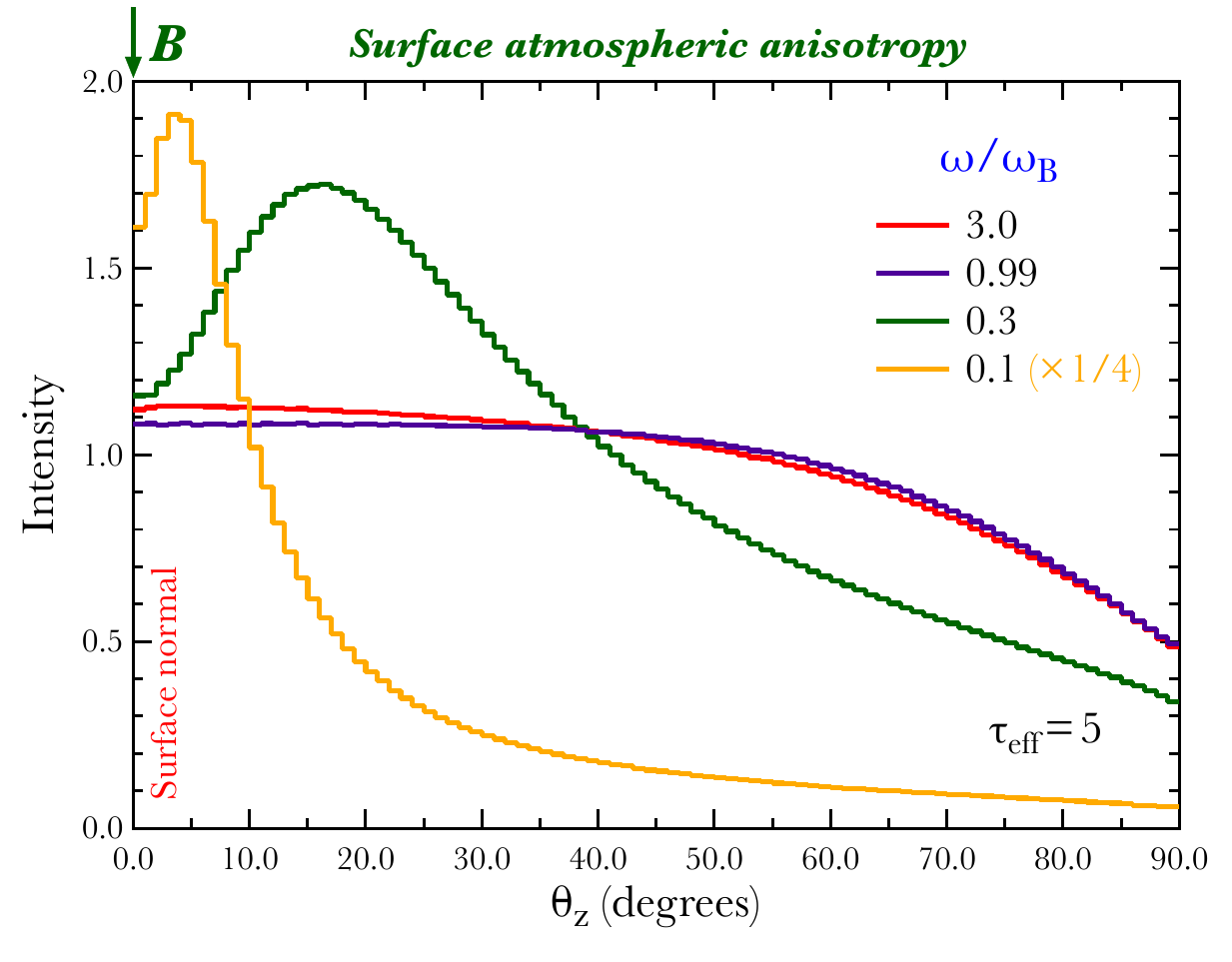}}
\caption{
\actionitem{Left}: 
Simulation geometry for transfer of photons through an atmospheric slab of Thomson 
optical depth \teq{\taut} with 
a normal direction along the \teq{z} axis (zenith), and magnetic field putatively in the \teq{x-z} plane,
at an angle \teq{\thetaB} to the local zenith.  All photons emerging from the upper boundary are recorded. 
\actionitem{Right}: 
Angular distributions for intensity \teq{I} emergent from the top of the 
atmospheric slab as functions of zenith angle \teq{\theta_z} for runs at the
magnetic pole (i.e., \teq{\thetaB=0^{\circ}}). 
Results are for frequency ratios \teq{\omega/\ecyc = 0.1, 0.3, 0.99, 3} as indicated, where 
\teq{\ecyc} is the cyclotron frequency.  The effective Thomson optical depth is fixed at \teq{\tau_{\rm eff}=5}. 
All histograms are for anisotropic and polarized injections at the slab base.  
Adapted from \cite{Hu-2022-ApJ}.
 \label{fig:geom_anis}}
\end{figure*}

X-ray polarimetry enables a new dimension in diagnostic space.  \IXPE\ has delivered 
key polarization results for four magnetars in its sensitive 2-8 keV band.
For 1RXS J1708-40, the pulse profile is single-peaked at 2-4 keV energies, 
and the polarization degree (PD) is anti-correlated with the intensity, while the
polarization angle (PA) was the same at all energies \citep{Zane-2023-ApJ}. 
This constant PA strongly contrasts the wide-ranging \IXPE~PA values 
seen in 4U 0142+61 between \teq{<4}keV (where the PD is 14\%) and \teq{>5.5}keV 
(PD is 42\%) energies \citep{Taverna-2022-Sci}.  The \teq{90^{\circ}} PA dichotomy 
is perhaps most naturally interpreted as the \teq{>5.5} keV signal emanating from 
a coronal region where resonant cyclotron scattering of surface photons occurs.  
SGR 1806-20 provided a further puzzle, with significant PD levels of 40\% only 
in the 4-5 keV window, and less than 5\% below 4 keV, leading to the suggestion 
\citep{Turolla-2023-ApJ}  that the thermal emission may originate in a 
condensed part of the star's surface.  

In this paper, the influences that both the anisotropy and polarization 
of radiation emergent from NS surfaces imprint upon their pulsed soft X-ray signals are addressed.
Our tool is a high fidelity, polarized radiative transfer code that is designed 
for treating neutron star atmospheres.  It is applied to extended surface regions, 
with the radiation trajectories in the magnetosphere being modified by general 
relativity.  Intensity and PD pulse profiles for a magnetar and a 
low-field pulsar are explored in two case studies that illustrate the codes efficacy.

\section{Modeling Atmospheric Radiative Transfer}
 \label{sec:simulate}

The modeling of neutron star atmospheres presented in this paper 
uses our versatile Monte Carlo simulation named {\sl MAGTHOMSCATT}   
\citep{Barchas-2021-MNRAS,Hu-2022-ApJ} that tracks 
{\bf polarized photon propagation and scattering} in strong fields threading atmospheric slabs.
The basic slab structure and transport schematic is shown at left in Fig.~\ref{fig:geom_anis}.  
The simulation models polarized magnetic Thomson opacity 
due to electrons in arbitrary (usually, but not necessarily uniform) magnetic field configurations 
for an optically thick, usually isothermal gas.   It applies to both 
pulsars and magnetars, its structure closely resembling that of
the simulation \cite{Whitney-1991-ApJS} developed for white dwarf applications.  
The cross section is the {\bf magnetic Thomson} one detailed in \cite{Barchas-2021-MNRAS}, 
which is highly anisotropic and possesses a strong dependence on photon polarization
and frequency \teq{\omega}; there is a profound resonance at the electron cyclotron 
frequency \teq{\ecyc = e B/m_ec}.  The code thus captures the intricate interplay 
between energy-dependent polarization and anisotropy.
The code's vertical stratification is in Thomson optical depth units, so it is simply 
adaptable to the atmospheric density gradients that are encountered in the hydrostatic 
structure. 

{\sl MAGTHOMSCATT}\ has been validated extensively \citep{Barchas-2021-MNRAS},
reproducing anisotropic angular distributions reported by \cite{Whitney-1991-ApJS} for
vertical fields ({\bf B} is along the zenith), and non-magnetic transport characteristics
detailed in the papers of \cite{Sunyaev-1980-AandA,Sunyaev-1985-AandA}.
The atmospheric transport simulation yields interesting non-uniform profiles 
for both zenith angle and surface azimuthal angle distributions when the field direction possesses 
large zenith angles \teq{\thetaB} at quasi-equatorial surface locales.   Examples of 
emergent radiation zenith angle distributions at the magnetic pole are provided 
in Fig.~\ref{fig:geom_anis}, right \citep[from Fig.~2 of ][]{Hu-2022-ApJ}.  
These highlight the differences in anisotropy signatures at different photon energies, 
particularly the profound collimation/beaming of radiation near {\bf B} when \teq{\omega \ll \ecyc}.

The simulation code also integrates over atmospheres covering large or complete portions 
of neutron star surfaces.  The magnetic field is specified as a dipole configuration embedded 
in the Schwarzschild metric \citep{Wasserman-1983-ApJ}.  In principle, the local temperature 
and brightness can be varied across the surface, although for the results presented here, 
they are held constant throughout the hot regions.  The propagation of the additive 
\teq{I,Q,U,V} Stokes parameters to infinity is non-dispersive and parallel-transported in the 
curved spacetime of the Schwarzschild metric.  This metric is an appropriate choice for 
slowly-rotating neutron stars, i.e. excluding millisecond pulsars. The key polarization 
measures are the  {\bf polarization degree} (PD) \teq{\Pi} and the {\bf polarization angle} 
(PA) \teq{\chi}, which are given by the familiar expressions
\begin{equation}
   \Pi \; =\; \sqrt{ \Bigl( \dover{Q}{I} \Bigr)^2 + \Bigl( \dover{U}{I} \Bigr)^2 }
   \quad ,\quad
   \chi \; =\; \dover{1}{2} \arctan \dover{U}{Q} \quad .
 \label{eq:polarization}
\end{equation}
Their values will depend on the tilt between the magnetic and 
rotation axes, and also the observer's viewing direction.
In tracking photon {\bf E}-field vectors throughout the atmospheric (scattering) and 
magnetospheric (propagation) transport, {\sl MAGTHOMSCATT}'s construction is
designed to permit routine extension to include photon {\bf E} rotation in
dispersive media (vacuum or plasma).  It is thus prepared for future 
inclusion of the influences of vacuum and plasma birefringence.

\section{Results}\label{sec:results}

To illustrate a variety of results from our simulations of extended atmospheres, 
the ensuing exposition focuses on two different isolated neutron stars.  The first is the magnetar 
1RXS~J1708-40, a bright pulsar of period 11sec,
at a distance of 3.8 kpc \citep{OK-2014-ApJS}\footnote{For a comprehensive summary of magnetar 
properties, see the McGill Online Magnetar Catalog 
at http://www.physics.mcgill.ca/~pulsar/magnetar/main.html}; it is amenable to X-ray polarimetry using 
\IXPE, due to its high spectral temperature, \teq{0.47}keV.  The second 
is the CCO pulsar PSR J0821-4300 in Puppis A (with period of 112 msec), which is of much 
lower magnetization; with its distance of 2.2 kpc and its lower effective surface temperature 
of \teq{0.21}keV \citep{Gotthelf-2009-ApJ}, it is faint enough to be difficult to measure its 
polarization using \IXPE.  The disparity of the spin-down fields of these pulsars 
underpins significant differences in their expected polarization degrees.
All simulations were performed for photon propagation to infinity through 
a Schwarzschild metric specified by a neutron star of mass \teq{\mns = 1.44 M_{\odot}}
and radius \teq{\rns = 10^6}cm.

\footnotetext{\textbf{Abbreviations:} CCO: central compact object}

\subsection{The Magnetar 1RXS~J1708-40}
 \label{sec:J1708}

The pulsar 1RXS~J1708-40 is a prominent magnetar  
with a rapid spin-down \citep{Israel-1999-ApJ} that leads to the inference of 
\teq{B_{p,\infty}\sin\alpha =9.3 \times 10^{14}}Gauss using a vacuum dipole 
electromagnetic torque on the star.  Here \teq{B_{p,\infty}} is its flat spacetime surface 
polar field strength, and \teq{\alpha} is the angle between \teq{\boldsymbol{\mu}} 
and the spin axis \teq{\boldsymbol{\Omega}}.  Its persistent pulsed emission includes 
quasi-thermal soft X rays below around 5 keV, with a nominal temperature of 0.45 keV 
\citep{Perna-2001-ApJ,Vigano-2013-MNRAS}.  This signal very likely emanates from 
the surface, and perhaps near its magnetic poles, with a pulse profile that 
varies along with its magnetospheric activity \citep{Younes-2020-ApJ}.  It also exhibits a steady, 
hard X-ray tail above 10 keV \citep{denHartog-2008-AandA} that is believed to be generated 
by the resonant inverse Compton scattering of the surface thermal photons 
by relativistic electrons in the magnetosphere \citep[e.g.,][]{BH-2007-ApandSS,Wadiasingh-2018-ApJ}.

\begin{figure*}[t]
\centerline{\includegraphics[width=250pt]{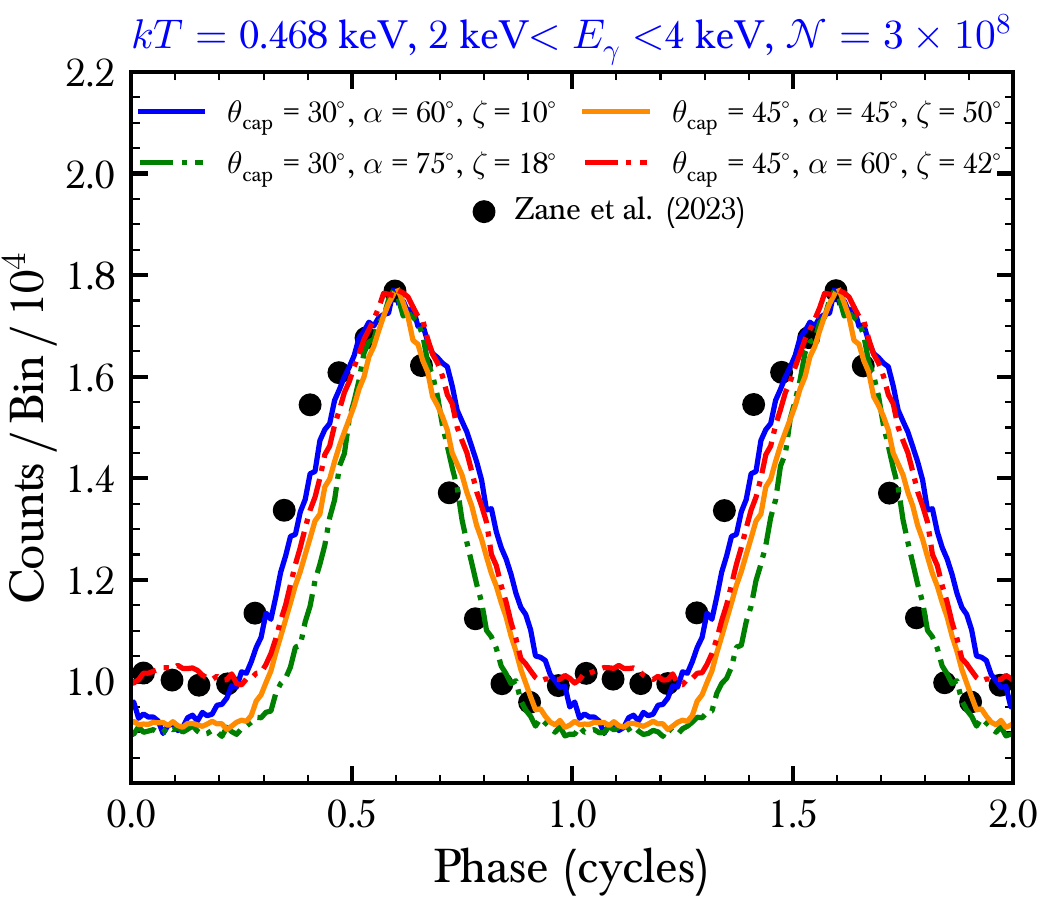}
\includegraphics[width=250pt]{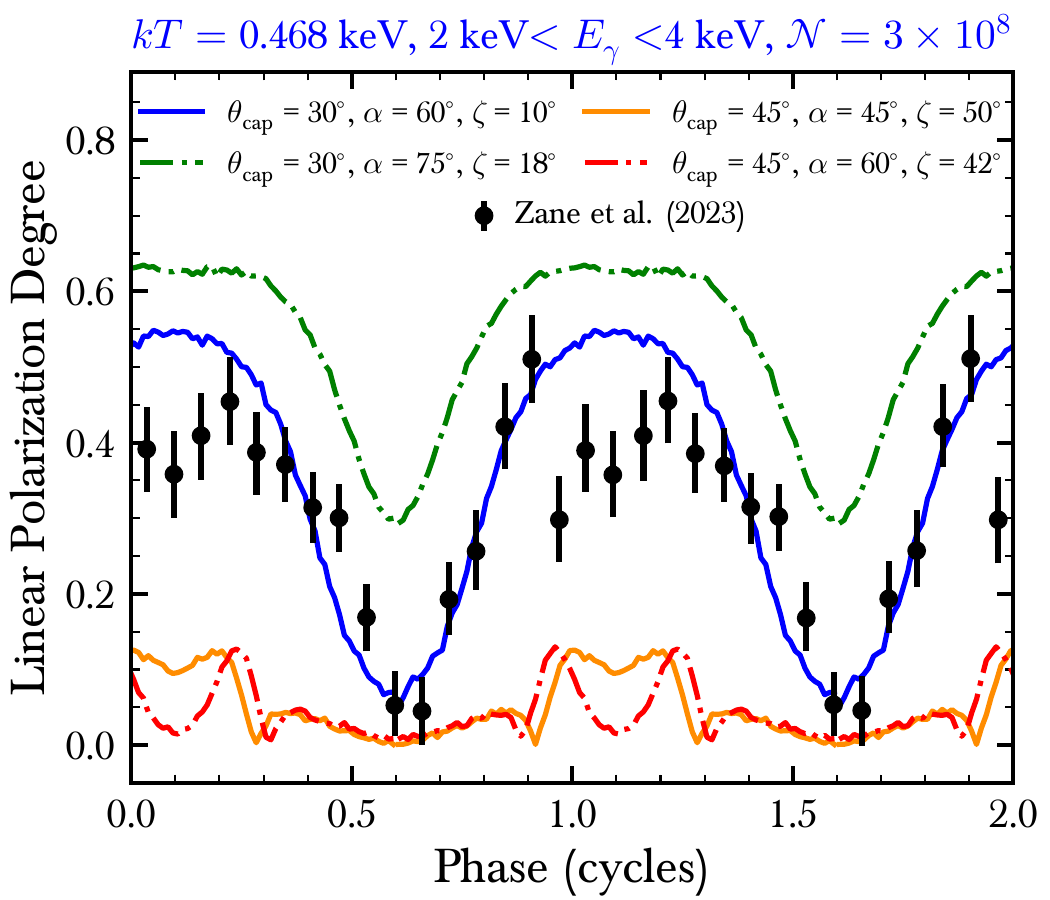}}
\caption{Simulated soft X-ray pulse profiles from {\sl MAGTHOMSCATT} for 
intensity \teq{I} \actionitem{(left)} and polarization degree (PD) \teq{\Pi} \actionitem{(right)} 
for two antipodal polar caps extending 
from the respective magnetic poles to colatitudes of \teq{\theta_{\rm cap} = 30^{\circ}, 45^{\circ}},
as labelled.  The points are \IXPE\ data from Fig.~4 of Zane et al. (2023) for the 
persistent 2-4 keV emission of the magnetar 1RXS J1708-40.  The different \teq{(\alpha, \zeta)} choices 
(see text) on the left all approximately describe the observed \IXPE\ light curve.  For the 
corresponding \teq{\Pi} at right, the models do not include the influences of 
vacuum birefringent propagation of light through the magnetosphere.
 \label{fig:I_PP_J1708}}
\end{figure*}

1RXS~J1708-40 has been well studied by a variety of X-ray telescopes over the 
years, and most recently by \IXPE.  The \IXPE\ polarization observations \citep{Zane-2023-ApJ} provide the 
focal point for the simulation runs that were completed in studying this magnetar and present here.  
Specifically, the 2-8 keV \IXPE\ spectrum can be fit by both two blackbody (BB+BB,
hot and cold) and blackbody plus power-law (BB+PL) superpositions.  
The intensity pulse profile is single-peaked in the softer 2-4 keV band, 
and moreover the polarization degree \teq{\Pi} is anti-correlated with the intensity.
These characteristics are displayed in Fig.~\ref{fig:I_PP_J1708}.  
The phase-averaged polarization angle \teq{\chi},
displayed in Fig.~1 of \cite{Zane-2023-ApJ}, was roughly the same at all energies
in the 2-8 keV band.  

To model this data, a number of simulations at different frequency 
ratios \teq{\omega /\ecyc} were performed.  This ensemble encompassed a 
range of frequencies appropriate to match the \teq{2-4}keV \IXPE\ dataset, and 
accommodate surface atmospheres with a polar field of \teq{9.3 \times 10^{14}}Gauss, 
as perceived by an observer at infinity.  They were weighted 
by a Planck spectrum of temperature \teq{kT\sim 0.47}keV.  The 
resulting photon energy ranges, as appreciated by an observer at infinity,
are listed in the panel headers in Fig.~\ref{fig:I_PP_J1708}.  These increase 
by a factor of \teq{1.32}, the inverse of the gravitational redshift factor 
\teq{\sqrt{ 1 - 2 G \mns /(\rns c^2)} \approx 0.76},
when measured in the local inertial frame at the stellar surface.
There are three main geometrical parameters, \teq{\alpha = \arccos \hat{\boldsymbol{\mu}}
\cdot \hat{\boldsymbol{\Omega}}}, the viewing angle \teq{\zeta} to the 
spin axis \teq{\boldsymbol{\Omega}}, and the colatitudinal extent \teq{\theta_{\rm cap}}
of each antipodal polar cap.  The solid angle-integrated emissivity is uniform across the
cap.  The goal is to constrain their values using the intensity 
and polarization degree light curves.

To do this, the data of all photons emitted from two identical antipodal surface 
polar caps to all directions on the sky is binned into observer angle cones of 
opening half-angle \teq{\zeta} and then sub-divided into bins of azimuthal angles 
around the spin axis \teq{\boldsymbol{\Omega}}, which then constitute different 
rotational pulse phases \teq{\Phi = \Omega t}.  The products are two-dimensional (2D)
``sky maps'' of intensity and polarization information in the \teq{\Omega t - \zeta}
plane, for each magnetic inclination angle \teq{\alpha}, which is 
increased in \teq{5^{\circ}} increments.  Examples of such sky maps are given 
in Figs.~4 and~5 of \cite{Hu-2022-ApJ}.  Horizontal cuts of such 2D maps then 
select light curves for a full range of values for \teq{\zeta}, again incremented in 
\teq{4^{\circ}} intervals.  

These light curves were then compared with the \IXPE\ intensity data, 
normalized so that the peaks coincide with those of the data. 
Using a visual inspection of a myriad of light curves (a \teq{\chi^2} statistical analysis 
was not possible due to the small error bars), 
the best ``fitting'' four pulse profiles are displayed in Fig.~\ref{fig:I_PP_J1708} (left)
for two polar cap sizes, \teq{\theta_{\rm cap} = 30^{\circ}, 45^{\circ}}.
These are clearly not good statistical fits, just general indications.  The 
symmetry of our dual polar cap constructions necessarily yields symmetric 
peaks and valleys in the light curves.  In contrast, the data possess a 
skewness that indicates surface non-uniformities in the emissivity; addressing 
such surface variations in detail is beyond the scope of this work.
The four models suggest a strong preference for an oblique rotator, with 
the magnetic inclination ranging from \teq{\alpha = 45^{\circ}} to 
\teq{\alpha = 75^{\circ}}.  Taken at face value, this would suggest a
bound \teq{9.6 \times 10^{14}}Gauss \teq{< B_{p,\infty} < 1.3 \times 10^{15}} Gauss.
Note that in principle, one should adjust the photon energy range 
for each inferred \teq{\alpha}, yet in practice, this protocol would make for only
a small difference in the model fitting parameter estimation.  The range 
of possible viewing angles \teq{\zeta} is also quite large.

The discrimination between these four best cases can be helped by inspection 
of the corresponding PD pulse profiles (which are not fits), displayed on the 
right of Fig.~\ref{fig:I_PP_J1708}.  Therein, the two cases with cap sizes 
\teq{\theta_{\rm cap} = 45^{\circ}} are strongly depolarized due to the 
large range of field directions spanned across the cap.  This depolarization 
trend is apparent when comparing the sky maps in Figs. 6 and 7 of \cite{Hu-2022-ApJ}.
The \teq{\alpha = 60^{\circ}, \zeta = 10^{\circ}} example in Fig.~\ref{fig:I_PP_J1708}
gives the best broad brush fit to the PD light curve.  However, this result 
was obtained with just the simple general relativistic parallel transport
of photon polarization vectors in the magnetosphere.  Therefore, our results
did not include the role of QED magnetic vacuum birefringence on the transport of 
polarizations to infinity; it preserves the polarization eigenstates and drive 
the net polarization degree much higher \citep{Heyl-2000-MNRAS}.  This influence is detailed 
in Dinh Thi et al. (in prep.), wherein PDs above 0.8 are routinely realized 
for 1RXS~J1708-40 using the same atmosphere output detailed here.  
Neither does our modeling include the competition between 
vacuum and plasma birefringence within the atmosphere, an influence 
that depolarizes surface signals overall \citep{Lai-Ho-2003-ApJ}, 
partially compensating the magnetospheric birefringence effects. 
Accordingly, future enhancements of our models including these 
influences will provide an improved 
picture of polarization predictions that would be better suited to \IXPE\
magnetar data comparison.

\begin{figure*}[t]
\centerline{\includegraphics[width=250pt]{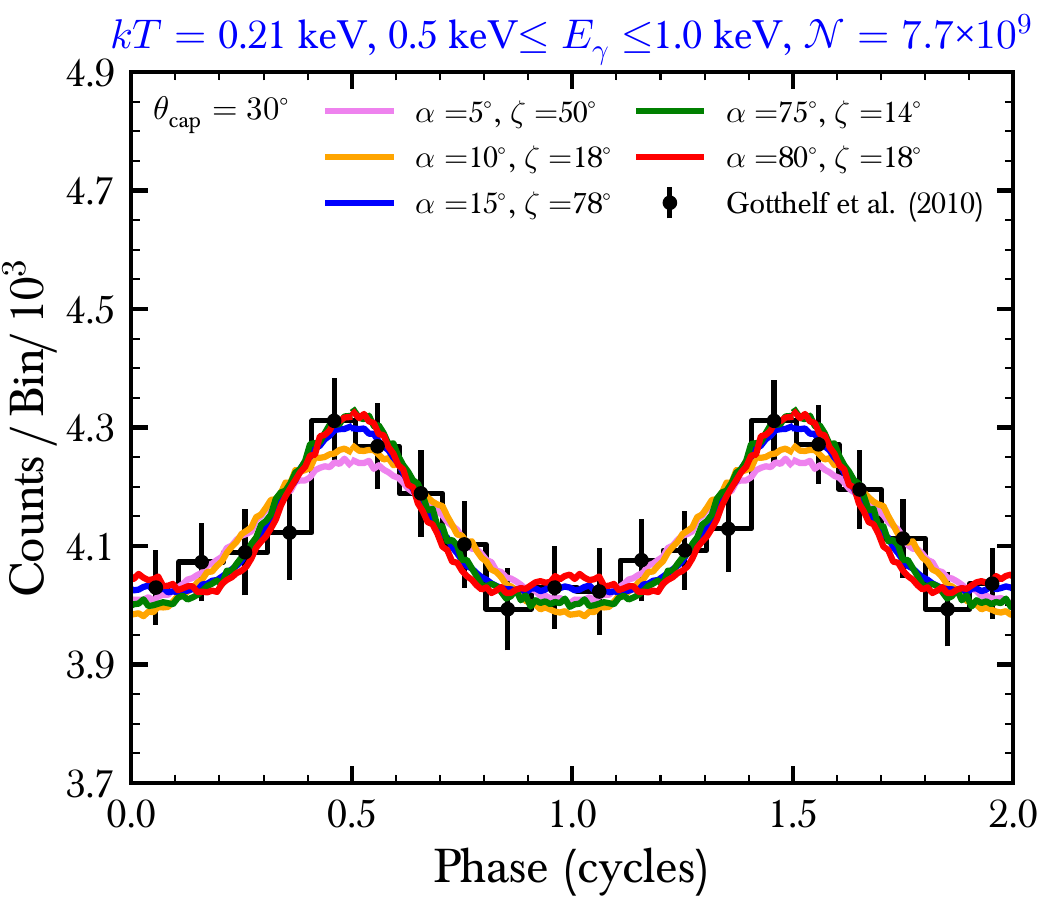}
\includegraphics[width=250pt]{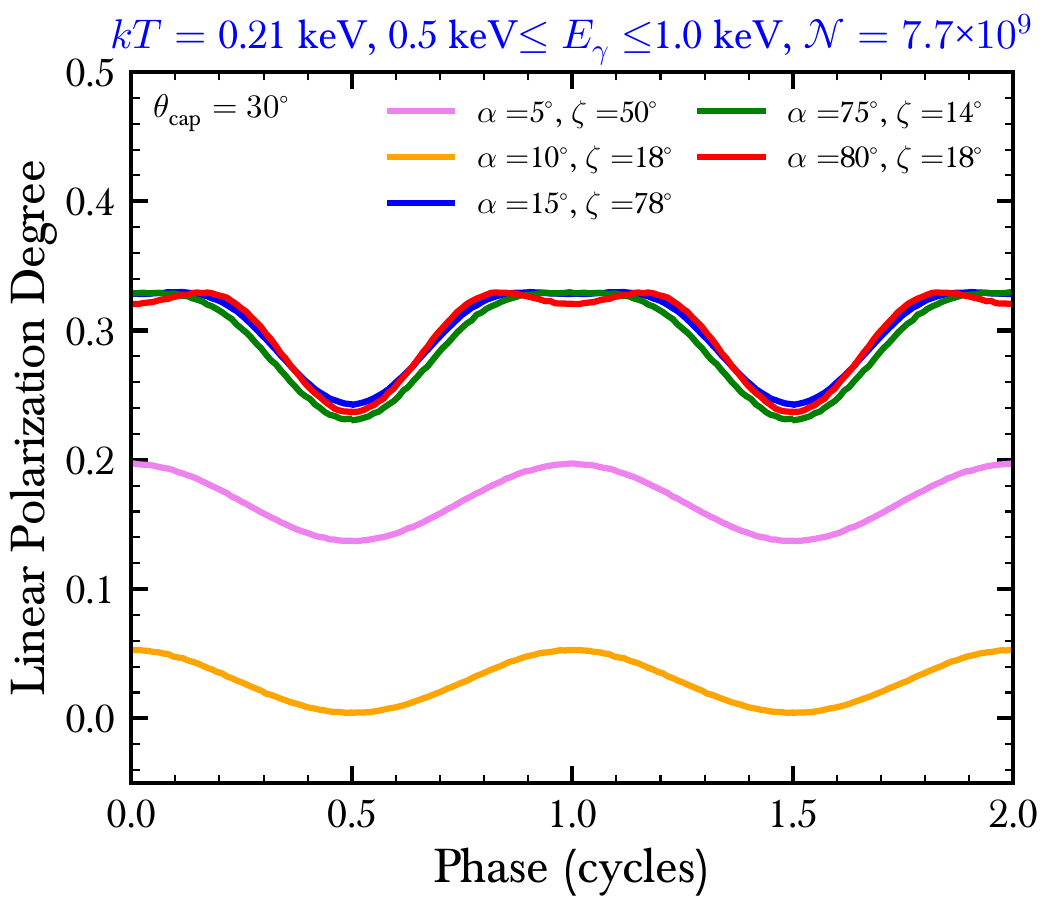}}
\caption{Simulated soft X-ray pulse profiles from {\sl MAGTHOMSCATT} for 
intensity \teq{I} \actionitem{(left)} and polarization degree (PD) \teq{\Pi} \actionitem{(right)} for the 
CCO pulsar J0821-4300 (see text).  The two antipodal polar caps extend 
from the respective magnetic poles to colatitudes of \teq{\theta_{\rm cap} = 30^{\circ}}.
The intensity data (left) are from the persistent 0.5-1 keV pulse profile 
displayed in Fig.~5 of \cite{Gotthelf-2010-ApJ}.  The different \teq{(\alpha, \zeta)} choices 
on the left all approximately describe the observed {\sl XMM-Newton} light curve.  
For the corresponding \teq{\Pi} at right, some of the models indicate the 
potential of polarimetry to help break parameter degeneracies.
 \label{fig:I_PP_J0821}}
\end{figure*}

\subsection{The CCO Pulsar PSR J0821-4300}
 \label{sec:J0821} 

Our second case study is for PSR J0821-4300 (more recently designated 
J0822-4300 due to its proper motion), an isolated neutron star 
of much lower magnetization.  Sometimes dubbed an ``anti-magnetar,''  its discovery did not
permit a measurable period derivative \teq{\dot P} at first \citep{Gotthelf-2009-ApJ}, 
though when eventually secured at \teq{{\dot P} = 9.3 \times 10^{-18}}, it led to 
the spin-down inference of \teq{B_{p,\infty} \sin\alpha =5.7\times 10^{10}}Gauss
for a vacuum rotator in flat spacetime \citep{Gotthelf-2013-ApJ}.  Another special
aspect to its character is its low pulsation amplitude (pulse fraction), around 8\%,
contrasting the large intensity modulation for 1RXS~J1708-40 discernible in 
Fig.~\ref{fig:I_PP_J1708}.  Interestingly,  the peak in the pulse profiles changed 
phase by \teq{180^{\circ}} between energies below 1.2 keV and those above this 
value \citep{Gotthelf-2009-ApJ}.  This feature was explained by \cite{Gotthelf-2010-ApJ} 
as signifying one polar cap being much hotter than the other.  

To explore how the light curves for this CCO pulsar can constrain the surface emission 
geometry, a similar protocol as in Sec.~\ref{sec:J1708} is adopted.  The intensity 
pulse profile that will be focused upon is that obtained from {\sl XMM-Newton} observations 
below 1 keV, and is displayed in Fig.~\ref{fig:I_PP_J0821}, left.  The geometry 
will be two identical temperature, antipodal polar caps extending from the magnetic poles out
to magnetic colatitudes \teq{\theta_{\rm cap}=30^{\circ}} distant.  This contrasts 
the set-up in \cite{Gotthelf-2010-ApJ}, wherein the two poles were of different 
sizes and temperatures.  Such an extension with our {\sl MAGTHOMSCATT} 
simulation is deferred to a future study, where its implications can be explored at 
length.  Here, a general sense of what the stellar geometric parameters 
\teq{\alpha = \arccos \hat{\boldsymbol{\mu}} \cdot \hat{\boldsymbol{\Omega}}}
(magnetic inclination) and viewing angle \teq{\zeta} are likely to be is posited.

Again, a number of simulations was performed at different frequency 
ratios \teq{\omega /\ecyc} that bracket the thermal photon energies 
pertaining to a temperature of around \teq{0.21}keV.  In the local 
inertial frame (LIF) at the surface, the gravitational redshift yields an increase 
in \teq{T_{\rm eff}} by \teq{1/\sqrt{ 1 - 2 G \mns /(\rns c^2)} \approx 1.32}.
These frequency results were then weighted with the appropriate Planck function.
The polar field strength as perceived at infinity was \teq{5.7 \times 10^{10}}Gauss,
which increases to \teq{8.5 \times 10^{10}}Gauss in the LIF at the surface, 
Again, the data of all photons emitted from a uniformly-emitting surface polar cap 
to all directions was binned into observer angle cones of opening half-angle 
\teq{\zeta} (in \teq{4^{\circ}} intervals) around the spin axis \teq{\boldsymbol{\Omega}}.
Thus 2D ``sky maps'' of intensity and polarization information in the \teq{\Omega t - \zeta}
plane were generated for each rotator inclination angle \teq{\alpha} that was stepped incrementally 
by \teq{5^{\circ}}.  Horizontal cuts of such 2D maps then generated the pulse profiles 
illustrated in Fig.~\ref{fig:I_PP_J0821}.

Of the multitude of resulting pulse profiles, five best fit models are illustrated 
in Fig.~\ref{fig:I_PP_J0821}.  Not all pulse profiles that were derived from 
the sky maps had pulse fractions this small.  These preferred models 
were determined using a \teq{\chi^2} statistics minimization approach 
when assuming Gaussian errors in the displayed PSR J0821-4300 counts 
(intensity) data.  Both small and large \teq{\alpha} were permissible, and 
similarly smaller and large \teq{\zeta} could be possible.  In the small 
\teq{\alpha = 5^{\circ},10^{\circ},15^{\circ}} examples, the remote polar cap 
is barely visible to an observer, so the peak at phase 0.5 is due to the 
proximate polar cap.  For the large \teq{\alpha} profiles, both caps are 
visible, yet the small observer inclination to the spin axis reduces the 
pulse fraction.  

Discriminating between these geometry parameters
using polarimetry is unfortunately not possible, since the low 
temperature renders PSR J0821-4300 a low flux target that proves difficult 
for \IXPE\ to successfully measure polarizations for.  Nevertheless, 
PD traces were also generated, on the right of Fig.~\ref{fig:I_PP_J0821},
to provide a general sense of what might be probed by a future, more sensitive
X-ray polarimeter.  Again, these are generated using parallel transport 
of photon electric field vectors in the Schwarzschild metric, from the 
stellar surface to infinity.

The polarization degrees overall are much lower than those for 1RXS~J1708-40.
This is principally because of the much lower magnetization.  The electron 
cyclotron energy in the local inertial frame is given by
\begin{equation}
   E_{\rm cyc} \; =\; \hbar \ecyc \; =\; \dover{e\hbar B}{m_ec}
   \; =\; 11.6\, \dover{B}{10^{12}\,\hbox{G}}\; \hbox{keV}\; .
 \label{eq:Ecyc}
\end{equation}
This sets up at \teq{0.66}keV for PSR J0821-4300 in the observer frame, right in the 
energy range of interest below 1 keV in Fig.~\ref{fig:I_PP_J0821}.  Thus the 
scattering in the atmosphere is around the cyclotron frequency, a domain in which the 
linear/circular polarization modes all have similar magnetic Thomson cross 
sections: see \cite{Barchas-2021-MNRAS}.  This leads to modest overall polarization
degrees, even for \teq{\theta_{\rm cap}=30^{\circ}}.  This contrasts the magnetar case, 
where \teq{\omega \ll \ecyc} always, and there is a strong dominance of 
the \teq{\parallel} linear mode (O-mode) over the \teq{\perp} one (X-mode), 
with minimal circular polarization.  This one-sidedness drives the net PD 
up to the large values evident here.

While the PD information separates off two cases as they have PD values 
inferior to 0.2, it is clear that three cases are essentially indistinguishable.
However, it was found that by isolating individual Stokes \teq{Q} and \teq{U} 
information via the polarization angle \teq{\chi} in Eq.~(\ref{eq:polarization}), 
the \teq{\alpha=15^{\circ}, \zeta = 78^{\circ}} case can be separated 
from the other two (not shown here).  Thus one has a path to constraining 
the \teq{\alpha ,\zeta} parameter space to around \teq{10^{\circ}} 
in a future advanced polarimetry era. Notwithstanding, it should again be 
mentioned that the PD levels do rise substantially when the effects of 
vacuum birefringence on the polarization modes throughout magnetospheric 
propagation is taken into account (Dinh Thi et al., in prep.).

\vspace{-5pt}
\section{Conclusion}
 \label{sec:conclude}
\vspace{-5pt}

This paper has presented sample results from our magnetic Thomson 
radiative transfer code {\sl MAGTHOMSCATT} that models atmospheres
of neutron stars of various magnetizations.  The polarized, anisotropic
signals are then propagated in curved spacetime through the magnetosphere to infinity.
The case studies presented here illustrate how our simulation data can be applied to both
a magnetar and an isolated, low-field pulsar to constrain key stellar 
geometry parameters, and the spatial extent of the emitting surface.
While light curve modeling efforts have been delivered before, our program brings 
greater depth to this approach through its precision simulation of 
atmospheric anisotropies and polarizations, and their coupling to magnetic colatitudes.
Thus, it is well positioned to interpret the X-ray polarization 
data being delivered by \IXPE.


\section*{Acknowledgments}

This work was performed with the support of the \fundingAgency{National Aeronautics and Space Administration}
under Grant Nos. \fundingNumber{80NSSC22K0853} and \fundingNumber{80NSSC24K0589}.

\bibliography{XMM_NS24_bdyh}

\end{document}